# What can we learn about cosmic structure from gravitational waves?

Joan M. Centrella

*Laboratory for High Energy Astrophysics, NASA Goddard Space Flight Center, Greenbelt, MD 20771 USA*

**Abstract.** Observations of low frequency gravitational waves by the space-based LISA mission will open a new observational window on the early universe and the emergence of structure. LISA will observe the dynamical coalescence of massive black hole binaries at high redshifts, giving an unprecedented look at the merger history of galaxies and the reionization epoch. LISA will also observe gravitational waves from the collapse of supermassive stars to form black holes, and will map the spacetime in the central regions of galaxy cusps at high precision.

## INTRODUCTION

Most of the information we have about the universe in general, and cosmic structure in particular, has been gleaned through detections of electromagnetic radiation. For example, observations with optical telescopes have revealed galaxy morphologies and clustering patterns; studies of galaxy spectra uncovered the universal recession as well as the presence of dark matter. Radio telescopes have shown a variety of active galaxies with vast cosmic jets emanating from their centers. Observations of the microwave sky revealed the 3K background radiation and the imprints of structure at early epochs. And high energy detectors unveiled a variety of exotic phenomena, ranging from hot gas in galaxy clusters to accretion disks around massive black holes (MBHs) at the centers of galaxies. This wealth of this electromagnetic information comes generally from excitations in gases in the outer regions of stars, within galaxies and clusers of galaxies, and in accretion flows around compact objects such as black holes.

As the $21^{st}$ century begins, a new observational window on the universe is being opened by detectors designed to measure gravitational waves from astrophysical sources. Gravitational radiation is produced through the bulk motion of massive objects such as black holes, and reveals their dynamical behavior directly. This article begins with a short introduction to gravitational radiation, and then encompasses a brief tour of several gravitational wave sources that can contribute to our understanding of cosmic structure.

## A GRAVITATIONAL RADIATION PRIMER

Gravitational waves are ripples in spacetime curvature that travel at the speed of light. They are typically generated by compact matter distributions that have time-changing quadrupole moments[1] such as binary systems and nonspherical collapses, and carry information about the bulk motion of the sources. Since gravitational waves couple very weakly to matter, they easily escape from the centers of galaxies or collapsed objects, bringing information about these deep, hidden regions. Gravitational waves carry both energy and momentum. For a binary system, this means that both the binary separation and the orbital period will decrease due to the emission of gravitational radiation. Such behavior has been observed in the binary pulsar PSR B1913+16, providing strong confirmation of the existence of gravitational radiation as predicted by general relativity [2].

The characteristic dimensionless gravitational wave amplitude for a source of mass $M$ located at a distance $r$ can be estimated as (e.g., [3])

$$h \sim \frac{G}{c^4}\frac{1}{r}\frac{d^2Q}{dt^2} \sim \frac{R_{\text{Schw}}}{r}\frac{v_{\text{ns}}^2}{c^2}, \quad R_{\text{Schw}} = \frac{2GM}{c^2}. \tag{1}$$

Here, $v_{\text{ns}}$ is the characteristic nonspherical velocity in the source, $Q$ is its (trace-free) quadrupole moment, and $R_{\text{Schw}}$ its Schwarzschild radius. Thus, the strongest sources have large masses moving with $v_{\text{ns}} \sim c$. For a stellar black hole system having mass $M = 10 M_\odot$, this gives $h \sim 10^{-16}$ at distance $r = 15\text{kpc}$, and $h \sim 10^{-21}$ at $r = 3000\text{Mpc}$. The gravitational waves from a MBH system with $M = 2.5 \times 10^6 M_\odot$ at a distance $r = 3000\text{Mpc}$ typically have amplitudes $h \sim 10^{-16}$.

The observed gravitational wave frequency of a source depends on its mass, physical scale, and redshift. For example, the nonspherical collapse of an object at redshift $z$ will produce a burst of gravitational waves with a characteristic frequency

$$f_{\text{burst}} \sim \frac{1}{2\pi\,t_{\text{dyn}}}\frac{1}{(1+z)} \sim (1.1 \times 10^{-2}\text{Hz})\frac{10^6 M_\odot}{(1+z)M}\left(\frac{R_{\text{Schw}}}{R}\right)^{3/2}, \tag{2}$$

where $t_{\text{dyn}} = \sqrt{GM/R^3}$ is the dynamical timescale. The nonspherical collapse of a star at $z \sim 0$ with mass $M \sim 10 M_\odot$ near its Schwarzschild radius $R \sim R_{\text{Schw}}$ can produce a burst of high frequency gravitational radiation at $f_{\text{burst}} \sim 1\text{kHZ}$. For the collapse of a supermassive star with $M \sim 10^6 M_\odot$ at $z \sim 2$, the radiation is produced at low frequency, $f_{\text{burst}} \sim 3.7\text{mHz}$.

A distorted black hole, such as might be formed from nonspherical collapse or the coalescence of two black holes or neutron stars, will emit gravitational waves as it settles down to a quiescent axisymmetric Kerr hole. These so-called ringdown waves have frequencies in the range [4, 5]

$$f_{\text{ring}} \sim (1.2 - 3.2) \times 10^{-2}\text{Hz}\left(\frac{10^6 M_\odot}{(1+z)M}\right), \tag{3}$$

where the lower end of the frequency range corresponds to a nonrotating Schwarzschild hole, and the upper end to a maximally rotating Kerr hole. Thus, a distorted stellar black hole with mass $M \sim 20 M_\odot$ in a nearby galaxy with $z \sim 0$ emits high frequency gravitational waves with $f_{\text{ring}} \sim 1\text{kHz}$. In contrast, a distorted MBH with mass $M \sim 2.5 \times 10^6 M_\odot$ rings down at low frequencies, with $f_{\text{ring}} \sim 4\text{mHz}$ at $z \sim 1$ and $f_{\text{ring}} \sim 1\text{mHz}$ at $z \sim 5$.

---

[1] Conservation of mass and momentum guarantee that there can be no monopolar or dipolar gravitational radiation, respectively. [1]

Another important class of gravitational wave sources consists of binary systems. The waves generated by a binary system whose components are on circular orbits and have comparable masses have frequency [6]

$$f_{\text{binary}} = \frac{2 f_{\text{orbital}}}{1+z} = \frac{1}{\pi} \frac{1}{(1+z)} \left( \frac{GM}{a^3} \right)^{1/2}, \qquad (4)$$

where $M$ is the total mass of the binary and $a$ is the separation of the binary components. As the gravitational waves are emitted, the binary separation $a$ decreases until it reaches $a \sim 3 R_{Schw}$, by which point the components have begun to merge. This gives a frequency range for these inspiral waves [7]

$$f_{\text{inspiral}} \lesssim 400 \text{Hz} \left( \frac{10 M_\odot}{(1+z) M} \right). \qquad (5)$$

Again we see that stellar black hole systems produce high frequency waves, so that $f_{\text{inspiral}} \lesssim 200\text{Hz}$ for a binary with total mass $M \sim 20 M_\odot$ at $z \sim 0$. MBH binaries emit waves at lower frequencies; a binary with total mass $M \sim 2.5 \times 10^6 M_\odot$ produces gravitational waves at frequencies $f_{\text{inspiral}} \lesssim 1\text{mHz}$ at $z \sim 1$ and $f_{\text{inspiral}} \lesssim 0.3\text{mHz}$ at $z \sim 5$.

## DETECTING GRAVITATIONAL WAVES

Broad-band gravitational wave detectors use laser interferometry to monitor the separation between test masses located at a distance $L$ from each other. When a gravitational wave impinges on a detector, the separation between the masses changes by $\Delta L$, resulting in a strain amplitude $h(t) = \Delta L/L$. Since gravitational wave detectors measure displacement $h \sim 1/r$ (see Eq. (1)), improving the sensitivity by a factor of 2 increases the regions of the universe from which sources can be detected by a factor of 8. Electromagnetic detectors generally measure energy flux $\sim 1/r^2$, so a similar increase in the observable volume would require improving the detector sensitivity by a factor of 4.

Ground-based kilometer-scale interferometers, such as LIGO, VIRGO, and GEO600, are sensitive to high frequency gravitational radiation in the range $10\text{Hz} \lesssim f_{\text{GW}} \lesssim 10^4 \text{Hz}$. For example, the LIGO (Laser Interferometric Gravitational-wave Observatory) project [8] has 2 interferometers with arm length $L = 4\text{km}$, one in Livingston, LA and and the other in Hanford, WA, and a third with $L = 2\text{km}$ at Hanford. LIGO has recently begun to take scientific data, and should be able to observe the dynamics of black holes with masses up to $\sim$ few $\times 10 M_\odot$ as well as other high frequency sources. The first generation interferometers are expected to reach their best strain sensitivities $h \sim 10^{-21}$ at frequencies near $f \sim 200\text{Hz}$. Second generation detectors can lower the broad-band sensitivity to $h \sim 8 \times 10^{-23}$ and achieve even better sensitivity with "tunable" narrow band techniques [9].

The planned space-based LISA (Laser Interferometric Space Antenna) mission [10] is a joint NASA/ESA collaboration designed to detect low frequency gravitational waves in the range $10^{-5}\text{Hz} \lesssim f_{\text{GW}} \lesssim 1\text{Hz}$. LISA consists of 3 spacecraft at the vertices of an equilateral triangle with arm lengths $L \simeq 5 \times 10^6 \text{km}$. The detector will follow 20° behind the Earth in its orbit about the Sun, and will observe gravitational waves from the dynamics of MBHs and other low frequency sources. LISA will reach its best strain sensitivity $h \sim 10^{-23}$ around frequency $f \sim 5\text{mHz}$ and should be capable of detecting MBH binaries to redshifts $z > 10$.

The timing of millisecond pulsars provides a means of probing the very low frequency gravitational wave band, $10^{-9}\text{Hz} \lesssim f \lesssim 10^{-7}\text{Hz}$. The rotation period of a millisecond pulsar is determined to high precision by measuring the arrival times of the pulses. Perturbations in the arrival times arise when the electromagnetic waves from the pulsar pass through distortions in spacetime caused by a stochastic background of gravitational waves [11], such as that from the ensemble of MBH binary coalescences across the universe. Current pulsar data sets reach a minimum characteristic strain $h \sim 4 \times 10^{-15}$ at frequencies $f \sim$ few nHz; observations with a planned Pulsar Timing Array could lower this sensitivity to $h \sim 10^{-15}$ [12].

## GRAVITATIONAL WAVES FROM MBH BINARIES

MBH binaries are expected to result from the merger of galaxies [13, 14] and, in hierarchical structure formation scenarios, the merger of halos containing seed black holes [15, 16]. A variety of astrophysical mechanisms have been proposed to bring a MBH binary from an initally wide orbital separation $a \gtrsim 1\text{kpc}$ to the regime in which gravitational radiation reaction will cause it to coalesce in less than a Hubble time. These include both dynamical friction from background stars [13, 17] and gas dynamical effects [18]. The time needed for a binary of total mass $M$ and separation $a$ to coalesce once gravitational radiation reaction becomes the dominant energy loss mechanism is [6]

$$t_{\text{GR}} = \frac{5}{64}\frac{c^5}{G^3}\frac{a^4}{M^3}, \tag{6}$$

assuming circular orbits and equal mass components. For example, a MBH binary of total mass $M \sim 2 \times 10^7 M_\odot$ starting from a separation $a \sim 10^{-2}\text{pc}$ will coalesce in a Hubble time $\sim 10^{10}$ years.

The final gravitational radiation-driven coalescence can be thought of as proceeding in 3 stages: an adiabatic inspiral, followed by a dynamical merger, and a final ringdown. As the binary evolves from inspiral through merger and ringdown, the gravitational wave frequency increases. As long as any accompanying accretion disks are dynamically negligible, black hole binaries coalescing through the emission of gravitational radiation are solutions to the vacuum (i.e., source-free) equations of general relativity. As such, their timescales are proportional to the total mass and, after time scaling, all other properties of the dynamics and waveforms depend only on ratios involving the masses and spins of the components. Thus, calcuations of the gravitational waveforms from all three stages of black hole binary coalescence can be easily scaled to apply to any masses and spins [3].

During the slow inspiral, the black holes are well-separated and spiral together adiabatically. The resulting gravitational waveform is a "chirp," which is a sinusoid that increases in frequency and amplitude as the orbital period shrinks; see Eq. (4). The gravitational waveforms from this inspiral phase are computed analytically using higher order post-Newtonian techniques in which the black holes are approximated as point masses [19]. These waveforms can be used as templates for data analysis algorithms based on matched filtering [20].

Typically, MBH binaries will be within the LISA frequency band for periods of roughly several months to years. If a sufficient number of cycles of the inspiral waveform are thus observed, both the so-called chirp mass $\mathcal{M}_c = (M_1 M_2)^{3/5}/(M_1 + M_2)^{1/5}$ and, with less accuracy, the reduced mass $m_{\rm red} = M_1 M_2/(M_1 + M_2)$ can be measured, as well as some information on the spins [21].

LISA's ability to detect a binary black hole inspiral depends on both the mass and redshift of the system. In particular, the gravitational wave frequencies of a binary of masses $M_1$ and $M_2$ at redshift $z$ are the same as those emitted by a binary local to the detector but having masses $(1+z)M_1$ and $(1+z)M_2$ [21]. Increasing the mass and/or the redshift lowers the frequency; see Eqs. (3) and (5). For binaries with mass $(1+z)M \sim 10^5 M_\odot$, most of the inspiral signal occurs in the frequency band in which LISA has very good sensitivity; this results in high signal-to-noise ratios (SNR) $\sim 10^2 - 10^3$ for such systems [21]. Overall, LISA should be able to detect the inspiral phase at SNR $\gtrsim 10$ for equal mass binaries of total mass $M \sim 10^3 M_\odot - 10^6 M_\odot$ out to $z \sim 9$, and higher mass inspirals $M \sim 10^7 M_\odot$ at lower redshifts $z \sim 1$ [21].

If MBH binaries proceed directly to the regime in which they coalesce under gravitational radiation reaction, recent calculations predict $\sim 10$ or more events per year will occur within LISA's frequency band [22, 23, 24]. Since many of these may originate at redshifts $z \gtrsim 7$, LISA can provide an outstanding probe of the high redshift universe, including the reionization epoch and the merger history of galaxies [24]. Observations in the nHz frequency range by pulsar timing arrays will detect the stochastic gravitational wave background from MBH binaries at lower redshifts, providing important constraints on that population [12].

The relatively slow inspiral is followed by the dynamical merger phase, in which the black holes plunge toward each other and merge into a single remnant, emitting a burst of gravitational waves. This stage is governed by strong nonlinear gravitational fields, and the resulting waveforms can only be calculated using full 3-D numerical relativity simulations [25, 26]. Typically, LISA will be able to observe gravitational wave bursts from this dynamical merger phase on timescales of roughly minutes to hours, providing an outstanding probe of a regime that is expected to be phenomenologically rich. For example, the merger of two black holes may cause a substantial change in the direction of the spin axis of the more massive hole [27]. The jets emanating from the center of an active galaxy are believed to be directed along the spin axis of the central MBH [28]. Thus, if such a hole suffered a change in spin direction as a result of merging with another MBH, the change in direction of the jet would cause a new radio lobe to be generated at some angle to the original one. This has been proposed as an explanation for the "winged" or "X-type" radio sources shown in Fig. 1 [29].

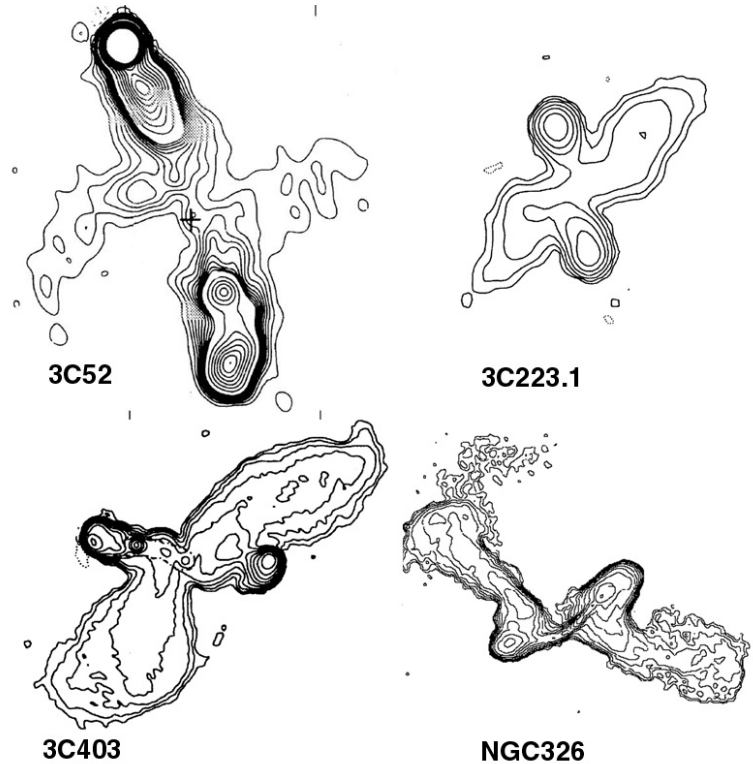

**FIGURE 1.** VLA radio observations of 4 radio galaxies showing the X-type morphology [30, 31, 32]. Reprinted with permission from Merritt, D. and Ekers, R. D., *Science*, **297**, 1310 (2002). Copyright 2002 American Association for the Advancement of Science.

After the dynamical merger is complete and a common event horizon forms, the distorted remnant will ring down to a quiescent Kerr hole through the emission of gravitational radiation in quasi-normal modes. The ringdown waveforms are known analytically from calculations of black hole perturbations, and take the form of damped sinusoids [4]. LISA will observe these bursts on time scales of typically minutes to hours. Detections at SNR $\gtrsim 10^2$ should be possible for masses $M \sim 10^5 - 10^7 M_\odot$ out to $z \sim 9$ for any value of the spin [21]. For $M \sim 10^8 M_\odot$ at $z \sim 1$, the ringdown occurs at the edge of LISA's low frequency sensitivity, and detection of the ringdown with large SNR requires large spin; c.f. Eq. (3) [33]. Observations of the gravitational waves from MBH ringdown will allow the identification of the mass and spin of the final Kerr black hole.

## SUPERMASSIVE STARS AND MBH FORMATION

The origin of MBHs is a topic of considerable interest for understanding the high redshift universe and the emergence of structure [34]. Various scenarios have been proposed, including the collapse of baryonic gas clouds (c.f. [35]) and dark matter halos [36].

MBHs may form hierarchically through the coalescence of smaller "seed" black holes. They may also form through direct collapse of large gas clumps; in fact, a recent study reveals the intriguing possibility that such MBHs may *form* in binaries [37].

A large gas cloud undergoing direct collapse may first form a rotating supermassive star as an intermediate state. Such an object is dominated by radiation pressure and is subject to a general relativistic radial instability leading to further collapse [6]. Recent numerical simulations have demonstrated that a uniformly rotating supermassive star collapses coherently to a black hole containing $\sim 90\%$ of the original mass; the rest of the gas forms a rotating disk outside the hole [38, 39]. A burst of gravitational waves from the collapse itself will be followed by ringdown radiation as the newly formed black hole settles into a quiescent state. For an interesting range of masses and redshifts, $f_{\text{burst}}$ and $f_{\text{ring}}$ lie within LISA's sensitive range; see Eqs. (2) and (3). If the supermassive star is differentially rotating, a non-axisymmetric instability could trigger bar formation before the catastrophic collapse; the rotation of this bar could generate quasi-periodic gravitational waves in the LISA band even if a MBH does not form [40].

## MAPPING MBH SPACETIMES

The final class of gravitational waves sources that we consider here consists of compact remnants of evolved stars – white dwarfs, neutron stars, or black holes – orbiting close to the central MBH in a galaxy [41, 42]. Compact remnants are expected to comprise a significant fraction of the total stellar population in the central cusp surrounding the MBH. If such an object is scattered into an orbit around the MBH, it will spiral in through the emission of gravitational radiation, eventually being captured by the MBH. If the remnant is on a tight eccentric trajectory, roughly the final year before capture will be spent executing $\sim 10^5 - 10^6$ orbits through the very strong field region near the event horizon of the MBH [43]. The gravitational radiation emitted during this stage can be detected by LISA at high accuracy [44, 45].

Since a stellar remnant of mass $\mu$ spiralling into a MBH of mass $M$ constitutes an extreme mass ratio binary, typically $\mu/M \lesssim 10^{-4}$, the motion of the remnant can be treated perturbatively in the MBH spacetime. The emission of gravitational radiation as the particle moves in the MBH potential causes the orbit to evolve. Technical issues arise in computing the effects of radiation reaction on the orbits and the accompanying waveforms [45]. Recent work on nonequatorial circular inspirals around a massive Kerr black hole shows that the gravitational waveforms are modulated by the effects of MBH spin and frame dragging, as shown in Fig. 2 [43]. The modulation is stronger at later times when the remnant is closer to the MBH event horizon. Complicated behavior also arises from eccentric orbits [46] and the spin of the remnant.

These trajectories and the resulting gravitational waveforms encode information about the multipolar structure of the gravitational potential of the central massive object. Measuring just three multipole moments can falsify whether this massive object is indeed a black hole [5]. LISA should be able to measure these extreme mass ratio waveforms out to a distance of a few Gpc if the remnant is a black hole with mass $\mu \sim 10 M_\odot$, and out to a few $\times 100$Mpc if the remnant is a white dwarf or neutron star [41, 42, 45]. Conservative estimates suggest that LISA will measure at least several such events per year, and possibly many more.

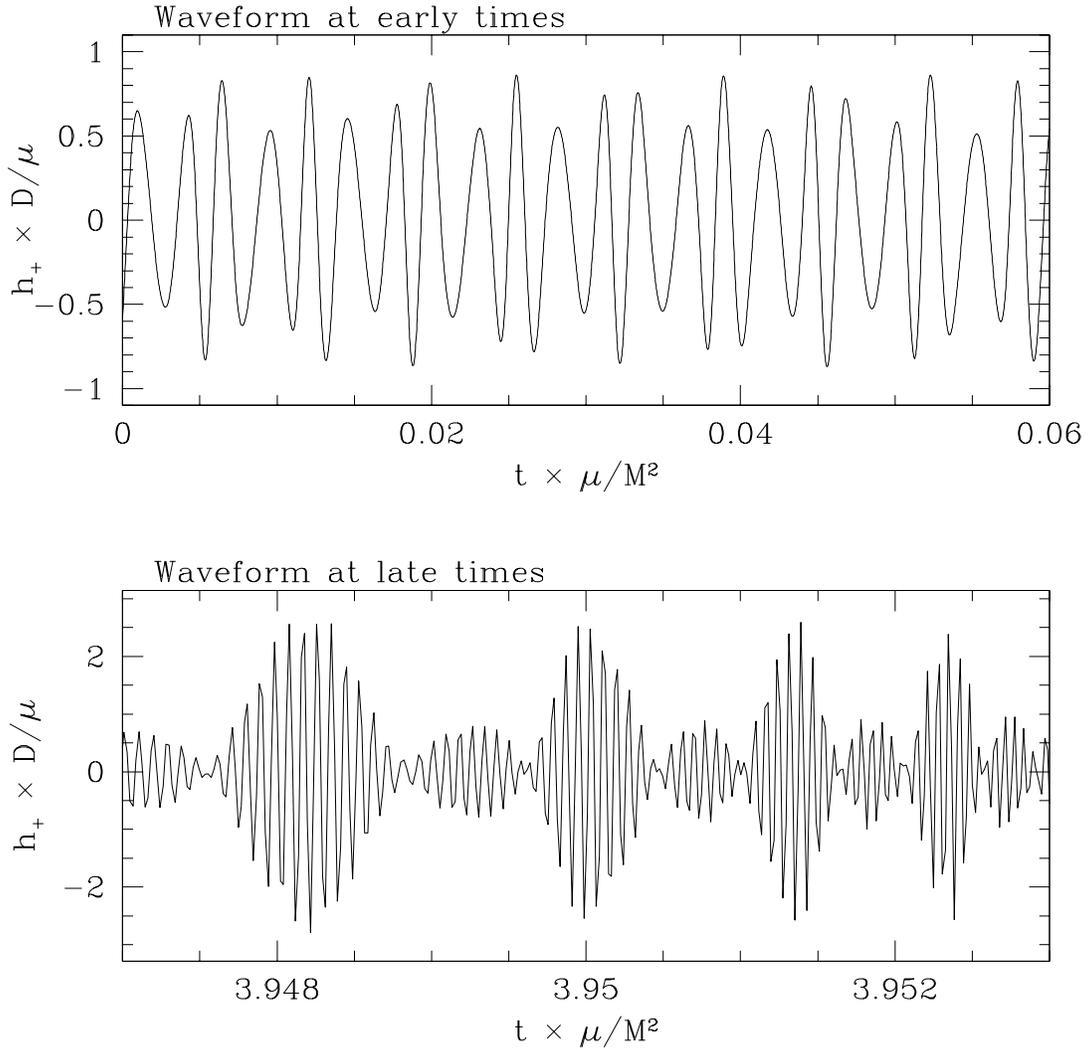

**FIGURE 2.** The gravitational waveform from an extreme mass ratio inspiral with $\mu/M = 10^{-4}$. The particle moves on a circular, nonequatorial trajectory around a central Kerr black hole which has nearly maximal spin, 0.998M. Reprinted from Ref. [43].

## SUMMARY

Gravitational wave detectors are poised to open a new observational window on the universe. Ground-based interferometers will observe the high frequency regime, $10\text{Hz} \lesssim f_{\text{GW}} \lesssim 10^4\text{Hz}$, including sources such as the dynamics of stellar black holes. The low frequency region of the spectrum, $10\text{Hz} \lesssim f_{\text{GW}} \lesssim 10^4\text{Hz}$, will be probed by the space-based LISA mission, and is especially important for understanding the emergence of structure in the universe. LISA will observe the coalescence of MBH binaries at high redshifts, yielding an unprecedented view of the merger history of galaxies and the reionization epoch. LISA will also be able to detect gravitational waves from the collapse of a supermassive star to a black hole, and will provide a high precision probe of the central regions of galaxy cusps. Pulsar timing arrays will observe the very low frequency regime, $10^{-9}\text{Hz} \lesssim f \lesssim 10^{-7}\text{Hz}$, which includes the stochastic gravitational wave background from MBH binaries at lower redshifts.

# ACKNOWLEDGMENTS

It is a pleasure to thank David Merritt for supplying Fig. 1; and Scott Hughes for stimulating discussions on MBH detection with LISA, and for supplying Fig. 2.

# REFERENCES

1. Misner, C. W., Thorne, K. S., and Wheeler, J. A., *Gravitation*, W. H. Freeman, New York (1973).
2. Weisberg, J. M., and Taylor, J. H., "The Relativistic Binary Pulsar B1913+16," in *Radio Pulsars*, edited by M. Bailes, D. J. Nice, and S. E. Thorsett, ASP Conference Series, 2003 (in press).
3. Thorne, K. S., "Probing Black Holes and Relativistic Stars with Gravitational Waves," in *Black Holes and Relativistic Stars*, edited by R. M. Wald, The University of Chicago Press, Chicago, 1992, pp. 41 - 78.
4. Echeverria, F., *Phys. Rev. D*, **40**, 3194 (1989).
5. Hughes, S. A., *Annals of Physics*, in press (2003).
6. Shapiro, S. L. and Teukolsky, S. A., *Black Holes, White Dwarfs, and Neutron Stars*, Wiley-Interscience, New York (1983).
7. Flanagan,É. É. and Hughes, S. A., *Phys. Rev. D*, **57**, 4535 (1998).
8. Barish, B. C., "First Generation Interferometers," in *Astrophysical Sources for Ground-Based Gravitational Wave Detectors*, edited by J. M. Centrella, AIP Conference Proceedings 575, American Institute of Physics, Melville, New York, 2001, pp. 3 - 14.
9. Fritschel, P., "The Second Generation LIGO Interferometers," in *Astrophysical Sources for Ground-Based Gravitational Wave Detectors*, edited by J. M. Centrella, AIP Conference Proceedings 575, American Institute of Physics, Melville, New York, 2001, pp. 15 - 23.
10. Bender, P., et al., *LISA, Pre-Phase A Report*, 2nd. edition, unpublished, available online at: http://lisa.jpl.nasa.gov/documents/ppa2-09.pdf (1998).
11. Phinney, S., astro-ph/0108028 (2001).
12. Jaffe, A. H., and Backer, D. C., astro-ph/0210148 (2002).
13. Begleman, M. C., Blandford, R. D., and Rees, M. J., *Nature*, **287**, 307 (1980).
14. Komossa, S., Burwitz, V., Hasinger, G., Predehl, P., Kaastra, J. S., and Ikebe, Y., *Ap. J*, in press (2003).
15. Islam, R. R., Taylor, J. E., and Silk, J., *Mon. Not. Roy. Astron. Soc.*, in press (2003).
16. Volonteri, M., Haardt, F., and Madau, P., *Ap. J.*, in press (2003).
17. Milosavljević, M. and Merritt, D., *Ap. J.*, **563**, 34 (2001).
18. Armitage, P. J. and Natarajan, P., *Ap. J.*, **567**, L9.
19. Will, C. M. and Wiseman, A. G., *Phys. Rev. D*, **54**, 4813 (1996).
20. Flanagan, É. É. and Hughes, S. A., *Phys. Rev. D*, **57**, 4566 (1998).
21. Hughes, S. A., *Mon. Not. Roy. Astron. Soc.*, **331**, 805 (2002).
22. Haehnelt, M. G., "Supermassive Black Holes as Sources for LISA," in *Laser Interferometer Space Antenna, Second International LISA Symposium on the Detection and Observation of Gravitational*

*Waves in Space*, edited by W. M. Folkner, AIP Conference Proceedings 456, American Institute of Physics, Woodbury, New York, 1998, pp. 45 - 49.
23. Menou, K., Haiman, Z., and Narayanan, V. K., *Ap. J.*, **558**, 535 (2001).
24. Wyithe, J. S. B. and Loeb, A., astro-ph/0211556 (2002).
25. Baker, J., Campanelli, M., Lousto, C. O., and Takahashi, R., *Phys. Rev. D* **65**, 124012 (2002).
26. Alcubierre, M., Bruegmann, B., Diener, P., Koppitz, M., Pollney, D., Seidel, E., and Takahashi, R., gr-qc/0206072 (2002).
27. Biermann, P. L., Chirvasa, M., Falcke, H., Markoff, S., and Zier, C., "Single and Binary Black Holes and their Active Environment,", in *High Energy Astrophysics from and for Space*, edited by N. Sanchez and H. de Vega, Proceedings of the 7eme Colloquium Cosmologie, Paris, in press (2002).
28. Begelman, M., Blandford, R., and Rees, M., *Rev. Mod. Phys.*, **56**, 255 (1984).
29. Merritt, D. and Ekers, R. D., *Science*, **297**, 1310 (2002).
30. Murgia, M., Parma, P., Ruiter, H. R., Bondi, M., Ekers, R. D., Fanti, R., and Fomalont, E. B., *Astron. Astrophys*, **380** 102 (2001).
31. Leahy, J. P. and Williams, A. G., *Mon. Not. R. Astron. Soc.*, **210**, 929 (1984).
32. Dennett-Thorpe, J., Bridle, A. H., Laing, R. A., and Scheuer, P. A. G., *Mon. Not. R. Astron. Soc.*, **304**, 27 (1999).
33. Hughes, S. A., private communication (2003).
34. Barkana, R. and Loeb, A., *Physics Reports*, **349**, 125 (2001).
35. Loeb, A. and Rasio, F., *Ap. J.*, **432**, 52 (1994).
36. Balberg, S. and Shapiro, S. L., *Phys. Rev. Lett.*, **88**, 101301, (2002).
37. Bromm, V. and Loeb, A., astro-ph/0212400 (2002).
38. Saijo, M., Baumgarte, T. W., Shapiro, S. L., and Shibata, M., *Ap. J.*, **569**, 349 (2002).
39. Shibata, M. and Shapiro, S. L., *Ap. J.*, **572**, L39 (2002).
40. New, K. C. B. and Shapiro, S. L., *Ap. J.*, **548**, 439 (2001).
41. Sigurdsson, S., *Class. Quantum Grav.*, **14**, 1425 (1997).
42. Sigurdsson, S. and Rees, M. J., *Mon. Not. R. Astron. Soc.*, **284**, 318 (1997).
43. Hughes, S. A., *Phys. Rev. D*, **64**, 064004 (2001).
44. Ryan, F. D., *Phys. Rev. D*, **56**, 1845 (1997).
45. Finn, L. S. and Thorne, K. S., *Phys. Rev. D*, **62**, 124021 (2000).
46. Glampedakis, K. and Kennefick, D., *Phys. Rev. D*, **66**, 044002 (2002).